\documentclass[aps, pre, twocolumn, showpacs, showkeys, superscriptaddress, amsmath, amssymb, a4paper]{revtex4-1} 
\usepackage{graphicx, dcolumn, bm, times, mathtools}

\usepackage{xcolor}

\begin{document}

\title{Relaxation with long-period oscillation in defect turbulence of planar nematic liquid crystals}

\author{Takayuki Narumi}
\email{narumi@ip.kyusan-u.ac.jp}
\affiliation{Faculty of Engineering, Kyushu Sangyo University, Fukuoka 813-8503, Japan}
\author{Yosuke Mikami}
\affiliation{Department of Applied Quantum Physics and Nuclear Engineering, Kyushu University, Fukuoka 819-0395, Japan}
\author{Tomoyuki Nagaya}
\affiliation{Department of Electrical and Electronic Engineering, Oita University, Oita 870-1192, Japan}
\author{Hirotaka Okabe}
\affiliation{Department of Applied Quantum Physics and Nuclear Engineering, Kyushu University, Fukuoka 819-0395, Japan}
\author{Kazuhiro Hara}
\affiliation{Department of Applied Quantum Physics and Nuclear Engineering, Kyushu University, Fukuoka 819-0395, Japan}
\author{Yoshiki Hidaka}
\email{hidaka@ap.kyushu-u.ac.jp}
\affiliation{Department of Applied Quantum Physics and Nuclear Engineering, Kyushu University, Fukuoka 819-0395, Japan}
\date{\today}

\begin{abstract} %
Through experiments, we studied defect turbulence, a type of spatiotemporal chaos in planar systems of nematic liquid crystals, to clarify the chaotic advection of weak turbulence.
In planar systems of large aspect ratio, structural relaxation which is characterized by the dynamic structure factor exhibits a long-period oscillation that is described well by a combination of a simple exponential relaxation and underdamped oscillation.
The simple relaxation arises as a result of the roll modulation while the damped oscillation is manifest in the repetitive gliding of defect pairs in a local area.
Each relaxation is derived analytically by the projection operator method that separates turbulent transport into a macroscopic contribution and fluctuations.
The analysis proposes that the two relaxations are not correlated.
The nonthermal fluctuations of defect turbulence are consequently separated into two independent Markov processes.
Our approach sheds light on diversity and universality from a unified viewpoint for weak turbulence.
\end{abstract}



\maketitle

\section{Introduction} \label{sec:introduction}
%
Convective systems have been studied as an example of chaos in dissipative systems \cite{Manneville1990}.
Some properties of weak turbulence in confined convective systems are characterized by chaos theory.
In spatially extended systems, weak turbulence exhibits spatiotemporal chaos.
This behavior is defined as a disorder state for which the correlation length is longer than the local order size much like convective roll pair \cite{Cross1993}.
Of current interest in spatially extended convective systems is how turbulence affects transport phenomena.
Transport in the vertical direction relates to the generation of convection, and the Nusselt number has been measured not only in convection but also in chaotic states \cite{Toth-Katona2003, Karimi2012}. 
On the other hand, it is also important that fluctuations of convectional structures bring horizontal transport, so-called chaotic advection \cite{Aref2014_arXiv}.
For example, the large-scale distribution of planktons on oceanic surface is a type of chaotic advection and significantly contributes in their ecosystem \cite{Abraham1998}.
Turbulent transport has been studied for developed turbulence appearing in strongly nonlinear regions, and concepts such as turbulent viscosity were established \cite{Monin1971}.
Developed turbulence has universal properties because strong nonlinearity breaks the characteristics of each system.
For chaotic advection in spatiotemporal chaos, in contrast, local order coexists with global disorder, and this makes it difficult to apply conventional treatments of fluid mechanics.
A common method needs to be established to explore universality, and also allows to accentuate diversity of each type of spatiotemporal chaos.
Here, analyzing structural relaxations by a statistical-physical method, we experimentally study chaotic advection of weak turbulence in a system of nematic liquid crystals.

Electroconvection in nematic liquid crystal systems has been studied as a suitable example of spatiotemporal chaos because one can prepare experimental systems with large degrees of freedom and a short response time \cite{Hidaka2014}.
Spatiotemporal chaos is generated near the threshold voltage for nematic electroconvection, and arises through the interaction between convective flow and the director that is the direction of preferred orientation of molecules.
In nematic liquid crystals, there are typically two types of alignment---planar and homeotropic, where alignment type responds to the boundary condition on the electrode substrates.
The types of alignment thus alter the symmetry of elctroconvective systems.

As to the symmetry, several types of spatiotemporal chaos emerge under the same nonequilibrium conditions.
With homeotropic alignment, the director is perpendicular to the electrodes and the continuous rotational symmetry on the electrode substrates exists.
As the applied voltage increases, the continuous rotational symmetry is broken spontaneously by the Fr\'eedericksz instability \cite{Freedericksz1933}, and thus the director behaves as a Nambu--Goldstone mode induced by symmetry breaking \cite{Hidaka2006}.
Above the threshold of electroconvection, the Nambu-Goldstone mode leads to a type of spatiotemporal chaos, called soft-mode turbulence (SMT) \cite{Kai1996, Hidaka1997_PRE}.
In contrast, with planar alignment, the nematic director is parallel to the electrodes.
Beyond a certain threshold of applied voltage, because the directivity forcibly breaks continuous rotational symmetry, a perfect stripe pattern, called normal roll (NR), appears.
At the second threshold, the axes of NR begin to fluctuate.
With fluctuations of NR, defects embedded in the stripe pattern nucleate in a spatially random manner \cite{Kai1989, Toth-Katona2003}, move across the electroconvective rolls (i.e., gliding motion), and then annihilate by the collision of them.
This is the type of spatiotemporal chaos investigated in the present study, called defect turbulence (DT).
Although DT and SMT appear under the same experimental conditions except for the boundary condition on the electrode substrates, their aspects are totally different, suggesting that the symmetry of system plays a significant role in spatiotemporal chaos.

Planar systems have been investigated traditionally since the discovery of electroconvection, and the study of DT has a long history \cite{Kai1989, Nasuno1992, Toth-Katona2003, Oikawa2004_JPSJ, Silvestri2009, Allegrini2009, Hidaka2015, Krekhov2015}.
There has been considerable research on the generation and dynamics of defects \cite{Kai1989, Toth-Katona2003}.
Since defects arise in association with nonlinear instability dynamics accompanied by phase fluctuations of periodical convection structures, the defect generation process partly contributes to the chaotic advection.
In addition, this can be also affected by structural fluctuations of the convective rolls.
Therefore, we should observe the convection structures for understanding the chaotic advection.

The temporal correlation of structural fluctuations, which is comparable to dynamic structure factor, is an useful quantity to investigate long-wavelength fluctuations in spatiotemporal chaos \cite{Ziman1979}.
In the previous works for SMT, the dynamic structure factor has been studied \cite{Nagaya2000, Narumi_SMTmem_2013}; especially in Ref.~\cite{Narumi_SMTmem_2013}, the dynamics was analyzed by so-called projection operator method that theoretically supplies hierarchical classification of dynamics in turbulence \cite{Mori2001}.
In this paper, the structural relaxation of DT is studied by observing the dynamic structure factor, and is analyzed by the projection operator method to clarify how symmetry affects transport phenomena in nonequilibrium open systems.
Further, as the method can be employed regardless of types of weak turbulence, we discuss diversity and universality from a unified viewpoint.

\section{Experiment}  \label{sec:experiment}
We researched the 2D pattern dynamics of DT observed in the planer alignment of nematic liquid crystals.
This study follows a standard setup \cite{Oikawa2004_PRE}.

The space between two parallel glass plates, spaced 50~$\mu$m apart, was filled with a nematic liquid crystal, MBBA (4-methoxy-benzilidene-4-n-butyl-aniline).
The plate surfaces were coated with transparent electrodes of size 1~cm $\times$ 1~cm, made of indium tin oxide.
To obtain planer alignment, the surfaces were covered by the surfactant, polyvinyl alcohol, and rubbed in one direction.
Denoting the dielectric constant parallel and perpendicular to the director by $\epsilon_{\parallel}$ and $\epsilon_{\perp}$, the dielectric constant anisotropy $\epsilon_{a}=\epsilon_{\parallel} - \epsilon_{\perp}$ was found to be negative.
An ac voltage, $V(t) = \sqrt{2}V\cos(2\pi f t)$, was applied to the sample.

One of the control parameters is the normalized voltage $\varepsilon = (V/V_c)^2-1$, where $V_c$ denotes the threshold voltage for nematic electroconvection.
DT appears above $\varepsilon\simeq 0.2$, and we performed experiments setting $\varepsilon = 0.2, 0.3, 0.4, 0.5$, and $0.6$. 
As the behavior for each $\varepsilon$ setting was similar qualitatively, we discuss only the results for $\varepsilon=0.4$ as being typical.
The temperature was regulated at 30.00 $\pm$ 0.05 $^{\circ}$C.
Before each sampling, we waited at the set $V$ to get a desired $\varepsilon$ value.
The waiting time was sufficiently long for systems to achieve steady state.
Another control parameter was the frequency $f$ of the ac voltage.
The frequency was set $f/f_{\text{c}} = 0.32$ with $f_\text{c} = 1100$~Hz, where $f_{\text{c}}$ is the critical frequency separating the conductive from the dielectric regime \cite{Dubois-Violette1971}.
The defect lattice did not appear under this condition \cite{Oikawa2004_PRE}. 

\begin{figure}[!t]
\includegraphics[width=0.95\linewidth]{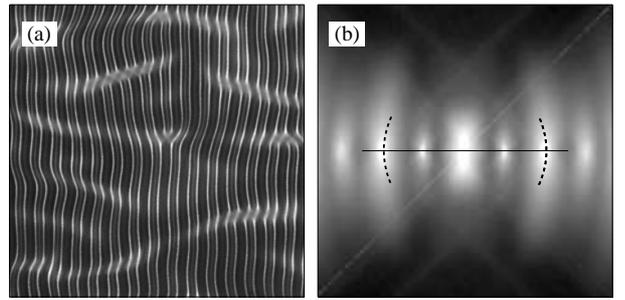}
\caption{%
	Images of DT at $\varepsilon= 0.4$.
	(a) A real-space snapshot of the transmitted light intensity $I(\bm{x},t)$.
	The brightness indicates the value of $I(\bm{x}, t)$.
	The horizontal direction corresponds to the $x$ direction.
	(b) Gray-scale magnitude of the spatial power spectrum $P_{\bm{k}}$ in $\bm{k}$ space.
	The brightness indicates the value of $P_{\bm{k}}$.
	The solid line represents the reference axis of the phase $\theta$ and the dashed arcs $|\bm{k}|=k_0$.
	\label{fig:experiment}
    }
\end{figure}

The patterns of transmitted light intensity $I(\bm{x},t)$ were observed under a microscope (ECLIPSE E600POL, Nikon Corporation, Tokyo, Japan) and captured by high-speed camera (Motion Scope M3, IDT Ltd., Hitchin, UK).
A snapshot showing clearly the anisotropy in DT at $\varepsilon=0.4$ is presented in Fig.~\ref{fig:experiment}(a).
At 10 frames per second, and 5000 images (i.e., 500 s of video) were taken in a single measurement.
The intensity at each pixel was digitized into 8-bit information.

To clarify the DT dynamics, we focused on the fluctuation $\Delta I(\bm{x},t) = I(\bm{x},t) - \left<I(\bm{x},t)\right>$ of the transmitted light intensity; here the angle brackets denote the long-time average in the steady state.
With the rubbing direction set as the $x$ direction and the axis of NRs aligned parallel to the $y$ axis, $\Delta I(\bm{x},t)$ was transformed into the spatial mode $u_{\bm{k}}(t)$:
\begin{equation}
	u_{\bm{k}}(t) = \int \Delta I(\bm{x},t)e^{i\bm{k}\cdot\bm{x}} d\bm{x},
	\label{eq:uk}
\end{equation}
where the range of integration is over the entire 2D domain.
The power spectrum $P_{\bm{k}} = \left<|u_{\bm{k}}(t)|^2\right>$ at $\varepsilon=0.4$ is illustrated in Fig.~\ref{fig:experiment}(b).
The reference length is the diameter $\lambda_0$ of a convective roll to observe the imaginary image, the focus of which exists near the roll axes \cite{Kondo1983, Rasenat1989}.
The corresponding wavenumber $k_0 = 2\pi/\lambda_0 = 0.294~\mu\text{m}^{-1}$ is the second peak of $P_{\bm{k}}$ in $\bm{k}$-space. 
We fixed the radius at $k_0$, and investigated the dependence of the phase $\theta$.
The stripes meander in the fluctuating NR, the degree of which is characterized by $\theta$; the origin $\theta = 0$ of the phase corresponds to NR.

\section{Result} \label{sec:result}

We calculated the normalized modal temporal correlation function to study the statistical properties of DT.
With the radius of $\bm{k}$ fixed at $k_0$, the function depends on the phase $\theta$ in $\bm{k}$-space.
Although the correlation can be a complex number in the absence of rotational symmetry, we focus on the real part of the modal correlation function
\begin{equation}
	U(\theta,\tau) = \text{Re}\left[\frac{ \left< u_{\bm{k}}(t+\tau) u_{-\bm{k}}(t)\right>}{P_{\bm{k}}}\right].
\end{equation}
The correlation of DT converges to zero in the long duration limit
\footnote{%
A few correlations converge to a finite value $C_{\infty} < 0$.
This is considered to have originated from the poor condition of the sample cells.
The modal correlations presented in this paper have been corrected so as to converge in the long-duration regime as $(U_{\text{experiment}}-C_{\infty})/(1-C_{\infty})$, where $U_{\text{experiment}}$ denotes the value obtained in experiments.
}.

\begin{figure}[!t]
\includegraphics[width=0.98\linewidth]{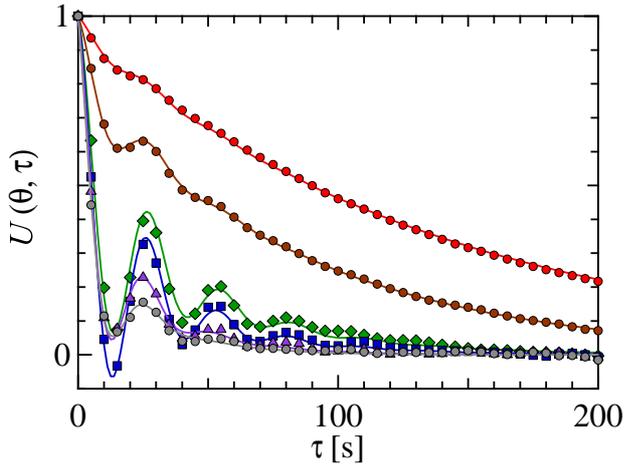}
\caption{(color online)
	The real part of the modal correlation functions at $\varepsilon=0.4$; $\theta = 0^{\circ}$ (red circle), $5^{\circ}$ (brown circle), $10^{\circ}$ (green diamond), $15^{\circ}$ (blue square), $20^{\circ}$ (purple triangle), and $25^{\circ}$ (gray circle), from top to bottom at the first peak ($\tau\simeq 25$ s).
	The solid lines are calculated from the expression \eqref{eq:relaxation}.
	The data in the plot were decimalized, whereas the values of the correlations were calculated at intervals of $0.1$~s.
	\label{fig:modalCF}
}
\end{figure}
The modal correlations represent a relaxation accompanied by the oscillation, the period of which is much longer than that for convection.
Figure~\ref{fig:modalCF} shows the modal correlation function of $\varepsilon=0.4$ for several $\theta$.
The relaxation $U(\theta, \tau)$ is described well by
\begin{equation}
	w U^{(\text{E})}(\theta,\tau) + (1-w)U^{(\text{D})}(\theta,\tau)
	\label{eq:relaxation}
\end{equation}
where $U^{(\text{E})}$ denotes a simple exponential relation
\begin{equation}
	U^{(\text{E})}(\theta,\tau) = \exp\left(-\frac{\tau}{\tau_{\text{E}}}\right)
	\label{eq:relaxation_E}
\end{equation}
and $U^{(\text{D})}$ a damped oscillation
\begin{equation}
	U^{(\text{D})}(\theta,\tau) = \exp\left(-\frac{\tau}{\tau_{\text{D}}}\right)
	\cos\left(\Omega_{\text{O}}\tau\right).
	\label{eq:relaxation_D}
\end{equation}
%
The parameter $w$ indicates the weight of the simple relaxation and the damping oscillation.
The simple relaxation is dominant when $w \simeq 1$.

\begin{figure}[!t]
\includegraphics[width=0.925\linewidth]{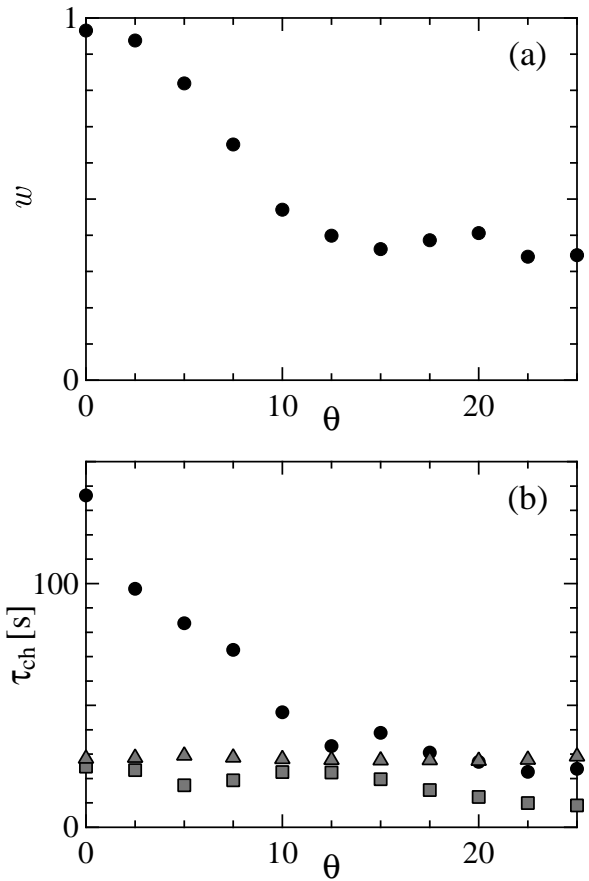}
\caption{%
	The angular dependence of the parameters in the expression \eqref{eq:relaxation} with Eqs.~\eqref{eq:relaxation_E} and \eqref{eq:relaxation_D}.
	(a) The weight parameter $w$.
	(b) The three characteristic times $\tau_{\text{ch}}$: $\tau_{\text{E}}$ (circle), $\tau_{\text{D}}$ (gray square), and $\tau_{\text{O}} = 2\pi/\Omega_{\text{O}}$ (gray triangle).
	\label{fig:w_and_tau}
}
\end{figure}
There are two types of $\theta$-dependence of the parameters.
The weight parameter decreases and converges as $\theta$ increases [Fig.~\ref{fig:w_and_tau}(a)].
The characteristic time $\tau_{\text{E}}$ for the simple relaxation shows a similar tendency.
In contrast, the characteristic times $\tau_{\text{D}}$ and $\tau_{\text{O}}=2\pi/\Omega_{\text{O}}$ of the damping oscillation are not affected as much by changes in $\theta$ [Fig.~\ref{fig:w_and_tau}(b)].

\section{Discussion} \label{sec:discussion}

The experimental results show that the modal temporal correlation oscillates over a macroscopic time scale. 
In this section, we first examine the origin of the exponential relaxation $U^{(\text{E})}$ and the damped oscillation $U^{(\text{D})}$, and then describe the relaxations analytically under two assumptions.

\subsection{Relaxation with long-period oscillation} \label{subsec:discussion1}

The origin of the exponential relaxation $U^{(\text{E})}$ is considered to be the relaxation of the roll modulation.
If the defects are negligible, by regarding DT as a fluctuating NR, $\Delta I(\bm{x},t)$ is simply expressed as
\begin{equation}
	\Delta I(\bm{x},t) = R_0 \exp [i (k_0 x + \alpha(y,t))] + \text{c.c.},
	\label{eq:phase}
\end{equation}
with $R_0$ a constant.
Such fluctuating NR relaxes to NR through diffusion of the phase $\alpha(y,t)$ in the $y$-direction.
The relaxation of $\alpha(y,t)$ modulation is described by $\exp(-t/\tau_{\alpha})$, where the relaxation time depends inversely on the square of the wavenumber, $\tau_{\alpha} \propto k_y^{-2}$.
This corresponds to the simple exponential relaxation of Eq.~\eqref{eq:relaxation_E}.

The $\theta$ dependence of $w$ and $\tau_{\text{E}}$ can be justified as follows.
The wavevector of the fluctuating NR is expressed as $\bm{k}_0+\nabla\alpha = (k_0, \partial \alpha/\partial y)$, where $\bm{k}_0 = (k_0,0)$ is the wavevector of NR.
As the deflection angle $\beta$ of rolls relative to the $y$-axis is given by
\begin{equation}
	\beta = \arctan\left(\frac{\partial \alpha / \partial y}{k_0}\right) \simeq \frac{\partial \alpha/\partial y}{k_0},
	\label{eq:beta}	
\end{equation}
the phase modulation with large $k_y$ includes large $\beta$.
The relaxation time thus shortens with increasing $\beta$.
As $\beta$ corresponds to $\theta$ in $\bm{k}$-space, the relaxation time $\tau_\mathrm{E}$ decreases with $\theta$ [Fig.~\ref{fig:w_and_tau}(b)].
Although this rough estimation cannot quantify $\tau_\mathrm{E}$, it helps in understanding why $\tau_\mathrm{E}$ depends on $\theta$.
Furthermore, the fluctuation consists of $\alpha$ modulations with various $k_y$.
The component for large $\beta$ is partially included only in the modulation with large $k_y$.
The contribution of $U^{(\text{E})}$ therefore decreases with $\theta$ [Fig.~\ref{fig:w_and_tau}(a)].
Based on the above discussion, the angle for which $w$ converges is interpreted as the limit below which the picture of phase diffusion holds for $U^{(\text{E})}$.
The torque under which the director is influenced from convection balances the twist at the limit angle.
Thus, the bending above the limit angle leads to defects.

The damped oscillation $U^{(\text{D})}$ can be understood from defect nucleation and its gliding motion in a local area
\footnote{%
	The climb motion does not occur because the characteristic length of the electroconvection keeps constant in our experiments.
}.
The instability of NR appears as the deflection of rolls by the interaction between the convection and the nematic director.
This leads to the Hopf instability found in the activator--inhibitor system consisting of the convective amplitudes and the twist angle of the nematic director \cite{Plaut1999}.
In a small system, for which the aspect ratio is $\mathcal{O}(1)$, the Hopf instability appears as a clear limit cycle called the angle-deflective oscillation \cite{Yang1986, Hidaka1992}.
In contrast, in the present system of the large aspect ratio, the instability of roll deflection immediately generates a defect pair at many different locations, and each defect glides in the reverse direction along the $x$-axis.
This obscures observation of the limit cycle %
\footnote{%
	Even in a large system, the oscillation appears clearly if the system is broken up into domains of abnormal rolls \cite{Oikawa2004_JPSJ}.
}.
Instead, we have directly found that the entire region of the system is separated into some areas containing a pair of defects; both defects return after reaching the lateral boundaries of this area, and the direction of the rolls reverts to $\beta \simeq 0$.
The repetition of the creation and annihilation develops into an oscillation with corresponding temporal correlation.
As expressed in Eq.~\eqref{eq:relaxation_D}, the oscillating amplitude decays because the sizes of the areas expand and contract with time
The characteristic times $\tau_\mathrm{D}$ and $\tau_\mathrm{O}$ of the damped oscillation are independent of $\theta$ because the local oscillation does not contribute to the wavenumber.

The temporal correlation of the injected power $P(t)$ (i.e., the Nusselt number) also oscillates with the longer period than that for convection \cite{Toth-Katona2003}.
The oscillation is not underdamped but rather persistent.
Although a more detailed study is needed to clarify the difference, we speculate that the difference is originated from that between a global variable $P(t)$ and a local one $I(\bm{x},t)$, where $P(t)$ detects the whole gliding motion of defects, and $I(\bm{x},t)$ does a part of the gliding motion.

We consider that the existence of an intermediate region is a universal feature of spatiotemporal chaos.
The above discussion suggests that DT has an enclosed region as the areas of the defects.
The scale is between the local order (i.e., the convection roll) and the system size.
The region can be regarded as the origin of the long correlation length of spatiotemporal chaos.
SMT contains such areas called a patch structure, in which the direction of the convective rolls is uniform \cite{Hidaka2006}.
The structure contributes strongly to the SMT dynamics enabling the dual structure to emerge that violates macroscopic time-reversal invariance \cite{Narumi_SMTmem_2013}.
Such regions have also been found to occur intermittently in not only nematic liquid crystals \cite{Oikawa2006, Takeuchi2007, Oikawa2008} but also other systems \cite{Kaneko1985, Pomeau1986, Chate1987, Ciliberto1988, Daviaud1990, Degen1996}.
In addition, a dynamic structure displaying dynamic heterogeneity was found in the slow dynamics of supercooled liquids near the glass transition, for which we proved similarity to SMT \cite{Nugroho_SMTtotal_2012}.
Seeking and investigating intermediate regions is an appropriate strategy in clarifying spatiotemporal chaos.

\subsection{Projection operator method} \label{subsec:discussion2}

Using the projection operator method \cite{Mori2001}, Eqs.~\eqref{eq:relaxation_E} and \eqref{eq:relaxation_D} are analytically explained from two assumptions about the dynamics of DT, specifically, (i) the absence of correlations between phase diffusion and instability of defect nucleation, and (ii) the Markovian nature of each nonthermal fluctuation.

The expression \eqref{eq:relaxation} suggests that mode $u^{(\text{E})}_{\bm{k}}(t)$ of the roll modulation is uncorrelated with mode $u^{(\text{D})}_{\bm{k}}(t)$ associated with instability of the defect nucleation.
We assumed that mode $u_{\bm{k}}(t)$ decomposes
\begin{equation}
	u_{\bm{k}}(t) = \sqrt{w}~u^{(\text{E})}_{\bm{k}}(t) + \sqrt{1-w}~u^{(\text{D})}_{\bm{k}}(t)
\end{equation}
with the conditions
\begin{equation}
	\left<u^{(\text{E})}_{\bm{k}}(t)u^{(\text{D})}_{-\bm{k}}(t^{\prime})\right> = \left<u^{(\text{D})}_{\bm{k}}(t)u^{(\text{E})}_{-\bm{k}}(t^{\prime})\right> = 0.
\end{equation}
The modal correlation is thus represented as in the expression \eqref{eq:relaxation} with
\begin{eqnarray}
	U^{(\text{E})}_{\bm{k}}(\tau) & = & \left<u^{(\text{E})}_{\bm{k}}(\tau)u^{(\text{E})}_{-\bm{k}}\right> / P_{\bm{k}}, \\
	U^{(\text{D})}_{\bm{k}}(\tau) & = & \left<u^{(\text{D})}_{\bm{k}}(\tau)u^{(\text{D})}_{-\bm{k}}\right> / P_{\bm{k}}.
\end{eqnarray}

The projection operator method is a useful tool to understand the relaxation dynamics of DT.
It separates chaotic or turbulent transport into a macroscopic contribution and a fluctuation.
We have applied it to SMT by treating turbulent disturbance as a nonthermal fluctuation \cite{Narumi_SMTmem_2013}.
In the formalism, the evolution of the vector $\bm{A}(t)$ of the macroscopic variables is analytically described by \cite{Mori1965}
\begin{equation}
    \frac{d}{dt}\bm{A}(t) = i\mathsf{K}\cdot\bm{A}(t)
- \int_{0}^{t}\mathsf{M}(t-t^{\prime})\cdot\bm{A}(t^{\prime}){\rm d}t^{\prime}
+\bm{R}(t),
    \label{eq:evolution_general}
\end{equation}
where $i\mathsf{K}$ denotes the frequency matrix, $\mathsf{M}(t)$ the memory matrix, and $\bm{R}(t)$ the fluctuating vector.
This equation is called the generalized Langevin equation.
The memory term depends on the transport coefficient through the fluctuation--dissipation relation.

One should first determine the macroscopic variables $\bm{A}(t)$ when using the projection operator method.
The chaotic dynamics of SMT has been successively elucidated simply by substituting $u_{\bm{k}}(t)$ for $\bm{A}(t)$.
The equation derived does not contain any oscillation terms.
For DT, we employed $u^{(\text{E})}_{\bm{k}}(t)$ as the macroscopic variable for the roll modulation, and the current $\dot{u}^{(\text{D})}_{\bm{k}}(t)$ in addition to $u^{(\text{D})}_{\bm{k}}(t)$ for the instability of the defects.
The evolution equations of each modal correlation are derived as
\begin{eqnarray}
    \dot{U}^{(\text{E})}_{\bm{k}}(\tau)
& = & - \int_{0}^{\tau} \Gamma^{(\text{E})}_{\bm{k}}(\tau-\tau^{\prime})
U^{(\text{E})}_{\bm{k}}(\tau^{\prime}) d\tau^{\prime}, \label{eq:EoM_cor_E} \\
    \ddot{U}^{(\text{D})}_{\bm{k}}(\tau)
& = & -\Omega_{\text{P}}^{2}U^{(\text{D})}_{\bm{k}}(\tau)
- \int_{0}^{\tau} \Gamma^{(\text{D})}_{\bm{k}}(\tau-\tau^{\prime})
\dot{U}^{(\text{D})}_{\bm{k}}(\tau^{\prime}) d\tau^{\prime}, \cr
    & & \label{eq:EoM_cor_D}
\end{eqnarray}
where $\Gamma^{(\text{E})}_{\bm{k}}(\tau)$ and $\Gamma^{(\text{D})}_{\bm{k}}(\tau)$
define memory functions, and $\Omega_{\text{P}}$ the angular frequency.
The derivation of Eq.~\eqref{eq:EoM_cor_D} is summarized in Appendix.

The forms of Eqs.~\eqref{eq:relaxation_E} and \eqref{eq:relaxation_D} relate to the Markov process.
Time evolutions are assumed to be Markovian at a macroscopic time scale, that is,
\begin{equation}
	\Gamma^{(\text{E})}_{\bm{k}}(\tau) = 2\gamma_{\text{E}} \delta(\tau)~,
	~~~\Gamma^{(\text{D})}_{\bm{k}}(\tau) = 2\gamma_{\text{D}} \delta(\tau)
	\label{eq:memory_markov}
\end{equation}
with $\gamma_{\text{E}}$ and $\gamma_{\text{D}}$ denoting the coefficients of friction.
Provided that $\gamma_{\text{D}} < 2\Omega_{\text{P}}$, Eqs.~\eqref{eq:EoM_cor_E} and \eqref{eq:EoM_cor_D} can be solved to obtain the functions defined in Eqs.~\eqref{eq:relaxation_E} and \eqref{eq:relaxation_D}.
The parameters are characterized as $\tau_{\text{E}} = 1/\gamma_{\text{E}}$, $\tau_{\text{D}} = 2/\gamma_{\text{D}}$ and $\Omega^2_{\text{O}}=\Omega_{\text{P}}^2 - \gamma^2_{\text{D}}/4$.

Tominaga et al. analyzed the chaotic Duffing oscillator using the projection operator method \cite{Tominaga2008}, and derived the same evolution equation as Eq.~\eqref{eq:EoM_cor_D} except for the external force term.
Because Eq.~\eqref{eq:EoM_cor_E} is caused by the phase diffusion in a spatially extended system, it does not appear in the Duffing system, which has no spatial degrees of freedom.

The form of the evolution equations, Eqs.~\eqref{eq:EoM_cor_E} and \eqref{eq:EoM_cor_D}, is similar to that in SMT \cite{Narumi_SMTmem_2013}.
However, the physical meaning of each of the memory functions is different because each macroscopic variable we employed is different.
The same macroscopic variables should be employed to compare DT with SMT in detail.

\section{Summary} \label{sec:summary}

In order to clarify the chaotic advection of weak turbulence, we have analyzed the patterns arising from the dynamics of DT using the dynamic structure factor.
Although the oscillation corresponding to the Hopf instability had been observed in systems of a small aspect ratio,
we have found relaxation to have long-period oscillations in systems of large aspect ratios.
The relaxation is represented by the superposition of the simple exponential relaxation $U^{(\text{E})}$ and the underdamped oscillation $U^{(\text{D})}$.
The weight parameter $w$ and the parameter $\tau_{\text{E}}$ in the simple exponential strongly depend on the roll deflection, whereas the parameters $\tau_{\text{D}}$ and $\Omega_{\text{O}}$ in the damped oscillation do not.
Extrapolating the relaxation mechanisms from the phase dependence of the parameters,
we conclude that the simple relaxation corresponds to the relaxation of the roll modulation and the damped oscillation originates from the reciprocal motion of the defect pairs in each local area.

The analysis by the projection operator method allows us to compare the chaotic advection in several types of weak turbulence from a unified viewpoint.
The relaxations of DT have been described analytically assuming no correlations between the two mechanisms, each fluctuation of which is Markovian.
They are different from the SMT dynamics, for which the nonthermal fluctuation is non-Markovian.
In contrast, DT and SMT are similar in having an intermediate region; the area of the defects in DT and the patch structure of SMT; this hierarchy may yield a clue to understanding the universality of spatiotemporal chaos.

\begin{acknowledgments}
	The authors gratefully acknowledge Hirotaka Tominaga for productive discussions.
	This work was supported by JSPS KAKENHI Grant Numbers JP24540408, JP15K05799.
\end{acknowledgments}

\appendix
\section{Derivation of Eq.~\eqref{eq:EoM_cor_D}} \label{sec:appendix}

We focus on macroscopic variables $\bm{A}(t)$ that satisfy the equations of motion
\begin{equation}
	\frac{{\rm d} \bm{A}(t)}{{\rm d}t} = i\mathcal{L}\bm{A}(t).
\end{equation}
where $i\mathcal{L}$ denotes the Liouville operator.
The projection operator $\mathcal{P}$ maps an arbitrary vector $\bm{Z}$ to a vector expressible in terms of $\bm{A} = \bm{A}(0)$;
\begin{equation}
    \mathcal{P}\bm{Z} = \left<\bm{Z}\bm{A}^{\dagger}\right>\cdot\mathsf{C}^{-1}\cdot \bm{A},
\end{equation}
where $\mathsf{C}$ denotes the correlation matrix $\left<\bm{A}\bm{A}^{\dagger}\right>$, the centered dot the inner product, the dagger the Hermitian conjugate, and $\left<\bm{X}\bm{Y}^{\dagger}\right>$ a matrix, e.g.,
\begin{equation}
	\bm{X} = \left(
	\begin{array}{c}
		x_{1} \\  x_{2}
	\end{array} \right)~~,~
	\bm{Y} = \left(
	\begin{array}{c}
		y_{1} \\  y_{2}
	\end{array} \right)
\end{equation}
gives
\begin{equation}
	\left<\bm{X}\bm{Y}^{\dagger}\right> =
	\left(
	\begin{array}{cc}
	\left<x_{1}y^{*}_{1}\right> & \left<x_{1}y^{*}_{2}\right> \\
	\left<x_{2}y^{*}_{1}\right> & \left<x_{2}y^{*}_{2}\right>
	\end{array}
	\right).
\end{equation}
The projection operator method yields the generalized Langevin equation \eqref{eq:evolution_general} for $\bm{A}(t)$.
Each term in the equation is analytically represented as \cite{Mori1965}
\begin{eqnarray}
	i\mathsf{K} & = &\left<\left(i\mathcal{L}\bm{A}\right)\bm{A}^{\dagger}\right>\cdot \mathsf{C}^{-1}, \\
	\bm{R}(t) & = & e^{\left(1-\mathcal{P}\right)i\mathcal{L}t}\left(1-\mathcal{P}\right)i\mathcal{L}\bm{A}, \\
	\mathsf{M}(t) & = & \left<\bm{R}(t)\bm{R}^{\dagger}\right>\cdot \mathsf{C}^{-1}.
\end{eqnarray}

As discussed in Section \ref{subsec:discussion2}, we employ mode $u^{(\text{D})}_{\bm{k}}(t)$ and its current $\dot{u}^{(\text{D})}_{\bm{k}}(t)$ to describe the instability of the defect nucleation.
The correlation matrix $\mathsf{C}$ is then represented as power spectra,
\begin{equation}
    \mathsf{C} = \left( \begin{array}{cc}
    \left<u^{(\text{D})}_{\bm{k}}u^{(\text{D})}_{-\bm{k}}\right> & \left<u^{(\text{D})}_{\bm{k}}\dot{u}^{(\text{D})}_{-\bm{k}}\right> \\
    \left<\dot{u}^{(\text{D})}_{\bm{k}}u^{(\text{D})}_{-\bm{k}}\right> & \left<\dot{u}^{(\text{D})}_{\bm{k}}\dot{u}^{(\text{D})}_{-\bm{k}}\right>
    \end{array} \right)
    =  \left(\begin{array}{cc}
    P^{(\text{D})}_{\bm{k}} & 0 \\
    0 & P^{(\text{D})\prime}_{\bm{k}}
    \end{array} \right),
\end{equation}
where $P_{k}^{(\text{D})\prime}$ is the power spectrum for the flux of the transmitted light intensity.
The frequency matrix $i\mathsf{K}$ reduces to
\begin{eqnarray}
	i\mathsf{K} & = & \left( \begin{array}{cc}
	\left<\left(i\mathcal{L}u^{(\text{D})}_{\bm{k}}\right)u^{(\text{D})}_{-\bm{k}}\right> & \left<\left(i\mathcal{L}u^{(\text{D})}_{\bm{k}}\right)\dot{u}^{(\text{D})}_{-\bm{k}}\right> \\
	\left<\left(i\mathcal{L}\dot{u}^{(\text{D})}_{\bm{k}}\right)u^{(\text{D})}_{-\bm{k}}\right> & \left<\left(i\mathcal{L}\dot{u}^{(\text{D})}_{\bm{k}}\right)\dot{u}^{(\text{D})}_{-\bm{k}}\right>
	\end{array} \right) \cdot \mathsf{C}^{-1} \cr
	& & \cr
	& = & \left(\begin{array}{cc}
		0 & 1 \\
		-P^{(\text{D})\prime}_{\bm{k}}/P^{(\text{D})}_{\bm{k}} & 0
	\end{array} \right).
	\label{eq:calc_frequency_term}
\end{eqnarray}
Note that the diagonal elements of $i\mathsf{K}$ are always zero, and this is the reason why the evolution equation of SMT in Ref.~\cite{Narumi_SMTmem_2013} does not contain any oscillation terms.
The fluctuation vector $\bm{R}$ at $t=0$ is found to
\begin{equation}
    \bm{R} = (1-\mathcal{P}) \left(\begin{array}{c}
		\dot{u}^{(\text{D})}_{\bm{k}} \\
		i\mathcal{L}\dot{u}^{(\text{D})}_{\bm{k}}
	\end{array} \right) = \left(\begin{array}{c}
		0 \\
	i\mathcal{L}\dot{u}^{(\text{D})}_{\bm{k}}
	+P^{\prime}_{\bm{k}}u^{(\text{D})}_{\bm{k}}/P_{\bm{k}}
	\end{array} \right).
\end{equation}
We denote the second element of $\bm{R}$ as
\begin{equation}
	R_{\bm{k}}
=i\mathcal{L}\dot{u}^{(\text{D})}_{\bm{k}}+P^{(\text{D})\prime}_{\bm{k}}u^{(\text{D})}_{\bm{k}}/P^{(\text{D})}_{\bm{k}}.
\end{equation}
The time evolution $R_{\bm{k}}(t)=e^{(1-\mathcal{P})i\mathcal{L}t}R_{\bm{k}}$ is called the fluctuating term.
Note that $\left<R_{\bm{k}}(t)u_{-\bm{k}}\right> = 0$.
The memory matrix $\mathsf{M}(t)$ then reduces to
\begin{equation}
    \mathsf{M}(t) = \left(\begin{array}{cc}
        0 & 0 \\ 0 &
        \left<R_{\bm{k}}(t)R_{-\bm{k}}\right>/P^{(\text{D})\prime}_{\bm{k}}
    \end{array}\right).
\end{equation}
Thus the equation of motion for $u^{(\text{D})}_{\bm{k}}(t)$ is
\begin{equation}
    \ddot{u}^{(\text{D})}_{\bm{k}}(t) = -\Omega^{2}_{\text{P}}u^{(\text{D})}_{\bm{k}}(t) - \int_{0}^{t}\Gamma^{(\text{D})}_{\bm{k}}(t-t^{\prime})\dot{u}^{(\text{D})}_{\bm{k}}(t^{\prime}) dt^{\prime} + R_{\bm{k}}(t).
    \label{eq:EoM_u}
\end{equation}
where $\Omega^2_{\text{P}} = P^{(\text{D})\prime}_{\bm{k}} / P^{(\text{D})}_{\bm{k}}$ denotes the angular frequency, and $\Gamma^{(\text{D})}_k(t)$ the memory function
\begin{equation}
    \Gamma^{(\text{D})}_k(t) = \left<R_{\bm{k}}(t)R_{-\bm{k}}\right>/P^{(\text{D})\prime}_{\bm{k}}.
\end{equation}
Multiplying Eq.~\eqref{eq:EoM_u} by $u_{-\bm{k}}$ and taking the average, one obtains Eq.~\eqref{eq:EoM_cor_D}.

%

%

\end{document}